\documentclass[a4paper,11pt]{article}
\pdfoutput=1
\usepackage{amssymb}
\usepackage{epsfig}
\usepackage{graphicx}

\makeatletter

\@addtoreset{equation}{section}
\makeatother
\def\be{\begin{equation}}
\def\ee{\end{equation}}
\def\bea{\begin{eqnarray}}
\def\eea{\end{eqnarray}}
\def\({\left(}
\def\){\right)}
\def\<{\left<}
\def\>{\right>}

\def\tr{{\mbox{tr}}}
\def\be{\begin{equation}}
\def\ee{\end{equation}}
\def\bea{\begin{eqnarray}}
\def\eea{\end{eqnarray}}
\def\ben{\begin{eqnarray}}
\def\een{\end{eqnarray}}
\def\({\left(}
\def\){\right)}
\def\<{\left<}
\def\>{\right>}

\def\[{\left[}
\def\]{\right]}

\def\+{\bar}
\def\mb{\mathbb}
\def\tr{{\mbox{tr}}}

\def\D{{\cal{D}}}

\def\t{\tilde}

\def\N{{\cal{N}}}

\def\be{\begin{equation}}
\def\ee{\end{equation}}
\def\bea{\begin{eqnarray}}
\def\eea{\end{eqnarray}}

\def\({\left(}
\def\){\right)}
\def\<{\left<}
\def\>{\right>}

\def\tr{{\mbox{tr}}}
\def\be{\begin{equation}}
\def\ee{\end{equation}}
\def\bea{\begin{eqnarray*}}
\def\eea{\end{eqnarray*}}
\def\ben{\begin{eqnarray}}
\def\een{\end{eqnarray}}
\def\({\left(}
\def\){\right)}
\def\<{\left<}
\def\>{\right>}
\def\!{\right|}
\def\|{\left|}

\def\[{\left[}
\def\]{\right]}

\def\+{\bar}
\def\mb{\mathbb}
\def\tr{{\mbox{tr}}}

\def\D{{\cal{D}}}

\def\t{\widetilde}

\def\R{{\cal{R}}}

\def\N{{\cal{N}}}

\def\F{{\cal{F}}}

\def\La{{\cal{L}}}

\def\E{{\cal{E}}}

\def\F{{\cal{F}}}

\begin{document}
\setlength{\unitlength}{1mm}

\pagestyle{empty}
\vskip-10pt
\vskip-10pt
\hfill %{\tt hep-th/yymmnnn}
\begin{center}
\vskip 3truecm
{\large \bf
Euclidean quantum M5 brane theory on $S^1 \times S^5$}\\ 
\vskip 2truecm
{\large \bf
Andreas Gustavsson\footnote{agbrev@gmail.com}}\\
\vskip 1truecm
{\it  School of Physics, Korea Institute for Advanced Study, Seoul 130-012,Korea}\\
and\\
\it{Physics Department, University of Seoul,  Seoul 130-743 Korea}
\end{center}
\vskip 2truecm
{\abstract{We consider Euclidean quantum M5 brane theory on $S^1\times S^5$. Dimensional reduction along $S^1$ gives a 5d SYM on $S^5$. We derive this 5d SYM theory from a classical Lorentzian M5 brane Lagrangian on $S^1 \times S^5$, where $S^1$ is a timelike circle of radius $T$, by performing a Scherk-Schwarz reduction along $S^1$ followed by Wick rotation of $T$.}}

\vfill 
\vskip4pt
\eject
\pagestyle{plain}

\section{Introduction}
Lagrangians for 5d SYM theories on $S^5$ with off-shell $\N=1$ supersymmetry were obtained in \cite{Hosomichi:2012ek}. Subsequently these theories were used to compute the $S^5$ partition function in the large $N$ limit where the instanton sectors are suppressed, by using localization \cite{Minahan:2013jwa,Kallen:2012va,Kallen:2012cs}. In \cite{Lockhart:2012vp,Kim:2012qf,Imamura:2012bm} it was found that these partition functions can be expressed in terms of the triple sine function, with an extension to squashed $S^5$. 

The field content of $\N=1$ SYM on $S^5$ is a vector multiplet in the adjoint representation and some number of masssive hypermultiplets in any representation. If we have just one hypermultiplet in the adjoint representation, and tune the hypermultiplet mass to a certain critical value, then supersymmetry is enhanced to $\N=2$ \cite{Kim:2012ava} which is maximal supersymmetry with $16$ supercharges. In \cite{Kim:2012ava} a 5d maximally supersymmetric YM theory (MSYM) on $S^5$ was obtained by dimensionally reducing Euclidean M5 brane theory on $S^1\times S^5$. Due to lack of a Majorana spinor in Euclidean six dimensions, the theory was defined in terms of complex spinors. It was found that  the bosonic part of this 5d MSYM Lagrangian is not real. This was observed only for the non-Abelian generalization where there is a cubic scalar interaction term that is not real. 

In \cite{Minahan:2013jwa} a different approach is taken. The 5d SYM theory that corresponds to Euclidean M5 brane on $S^1\times S^5$ is guessed by matching the free energy computed from the SYM theory with a computation in AdS space. In this case the bosonic Lagrangian becomes real for an R symmetry group that is a subgroup of $SO(1,4)$ rather than a subgroup of $SO(5)$. But in this approach the relation to the M5 brane is not clear. 

In this paper we approach the problem in a third way. Instead of dimensionally reducing Euclidean M5 brane, we reduce Lorentzian M5 brane along a time-like circle \cite{Hull:2014cxa}. Wick rotation to Euclidean M5 brane is then carried out in the 5d SYM theory. Apart from these differences, our approach follows that of \cite{Kim:2012ava}. We will argue from the M5 brane picture that the hypermultiplet mass in the 5d SYM theory should be Wick rotated to go to the Euclidean M5 brane theory. We then find that the 5d SYM Lagrangian is real when R symmetry is a subgroup of $SO(1,4)$, which is in agreement with the proposal in \cite{Minahan:2013jwa}. We also find that we can not keep $\N=2$ supersymmetry manifest during this Wick rotation to the Euclidean M5 brane.

Lorentzian flat Abelian M5 brane on $\mb{R}^{1,5}$ with $SO(1,5)\times SO(5)$ Lorentz times R-symmetry has a classical field theory description in terms of the $(2,0)$ tensor multiplet with fermions that satisfy 11D Majorana condition and 6d Weyl condition. 

Lorentzian M5 brane can also be put on $\mb{R} \times S^5$ while preserving $32$ superconformal symmetries where time $t$ is along $\mb{R}$. If we compactify $\mb{R}$ this will break all superconformal symmetry since these superconformal symmetries have non-trivial time-dependences $e^{\pm i \frac{T}{2r}t}$. Here $r$ is the radius of $S^5$ and $T$ is the distance along the time axis that we traverse as $t$ traverses a $2\pi$ interval. If we gauge the $SO(5)$ R-symmetry by introducing an extra gauge field that we declare has a trivial superconformal variation, then a new possibility arises when we compactify time $t\sim t + 2\pi$ into a timelike circle. We can turn on a nontrivial holonomy
\bea
P \exp i \int_{S^1} dt A_t
\eea
around the $S^1$. Let us begin with trivial holonomy, $A_t=0$, and then perform a gauge transformation by an element $g(t)$ in $SO(2) \times SO(2) \subset SO(5)$ that depends nontrivially on $t$ and which does {\sl{not}} respect periodic boundary condition in $t$. This will bring $A_t$ into the flat connection
\bea
A_t &=& i g \partial_t g^{-1}
\eea
and it will turn on a nontrivial holonomy. This is not a gauge transformation in the usual sense that the holonomy is invariant, precisely because $g(t) \neq g(t+2\pi)$. The effect of this gauge transformation on the superconformal transformations can be traded for a transformation of the superconformal parameter $\epsilon \rightarrow g\epsilon$. (We explain these steps in detail in Appendix \ref{B}). Thus by choosing $g$ suitably, a possibility arises to cancel some of the $t$ dependence in the original $\epsilon$ parameter. However, we can not cancel the $t$-dependence for all the $32$ parameters. We can at most do this for $16$ of these parameters, and more generally we can do it for only $8$ parameters. 

Thus we can put M5 brane on $S^1\times S^5$ with $S^1$ a timelike circle, by turning on a nontrivial holonomy around $S^1$. This holonomy is physically observable. Upon dimensional reduction along $S^1$ the holonomy will turn into a mass parameter of the 5d $\N=1$ hypermultiplet on $S^5$. The conjecture in \cite{Douglas:2010iu,Lambert:2010iw} says that M5 brane theory is equivalent with the 5d SYM theory that one obtains by dimensional reduction along a circle. If this conjecture is correct, then perhaps all the 5d theories with various hypermultiplet masses, would be equivalent with the corresponding 6d theory with corresponding holonomies around $S^1$. Or perhaps we only have the weaker version that applies to the maximally supersymmetric case which is that with $16$ superconformal symmetries in 6d, that upon dimensional reduction become $16$ ordinary Poincare supersymmetries of the 5d SYM theory. To extend the test using AdS/CFT to theories with $8$ supersymmetries, it seems that one would need to understand how to map a generic holonomy around the $S^1$ into the AdS side, and perhaps the conjecture does not apply to this case with less amount of supersymmetry despite the field content is the same as that of the maximally supersymmetric case.

%\subsection{A toy model}
We may understand how to pass from Lorentzian to Euclidean quantum theory by considering a toy model in $1+1$ dimensions with the Lagrangian 
\bea
L &=& \frac{1}{2} \int dx \((\partial_t\phi)^2 - (\partial_x \phi)^2\)
\eea
The conjugate momentum is $\pi = \partial_t\phi$ and the Hamiltonian is $H = \int dx \pi \partial_t\phi - L$. We may then take the Hamiltonian 
\ben
H &=& \int dx \(\frac{1}{2} \pi^2 + \frac{1}{2}(\partial_x \phi)^2\)\label{haminapp}
\een
together with the equal time commutation relation
\bea
[\phi(t,x),\pi(t,x')] &=& i \delta(x-x')
\eea
as the definition of the quantum theory. 

The amplitude for the transition from an initial state $\|\phi_i\>$ at time $t=0$ to a final state $\|\phi_f\>$ at time $t=2\pi T$ is computed as
\bea
A(\phi_f,\phi_i) &=& \<\phi_f\! e^{-2\pi iTH}\|\phi_i\>
\eea
We obtain the corresponding amplitude in the Euclidean theory as\footnote{I would like to thank Kimyeong Lee for introducing to me this definition of Euclidean quantum theory in the context of the Euclidean M5 brane.}
\bea
A_E(\phi_f,\phi_i) &=&  \<\phi_f\! e^{-2\pi RH}\|\phi_i\>
\eea
where $2\pi R$ is the real-valued Euclidean time interval. We use the same Hamiltonian and the same canonical commutation relations as in the Lorentzian theory. To pass to the Euclidean theory we only need to Wick rotate $T$ by first replacing $T=-iR$ and then rotating $R$ by 90 degrees into the real axis. We show how to derive the Euclidean action and the associated Wick rotation of time $t = -it_E$ from this Wick rotated amplitude in the Appendix \ref{Q}. 

We would now like to change a convention. Instead of defining the Hamiltonian so that it evolves $t$ from $t=0$ to $t=2\pi T$, we will in this paper define the Hamiltonian so that it evolves the parameter $t$ from $t=0$ to $t=2\pi$. This can be achieved by using the metric 
\ben
ds^2 &=& -T^2 dt^2 + dx^2\label{nmetr}
\een 
The Lorentz invariant Lagrangian is now given by 
\bea
L &=& \int dx \frac{T}{2} \(\frac{1}{T^2}(\partial_t\phi)^2 - (\partial_x \phi)^2\)
\eea
The insertions of the various powers of $T$ come from the usual way of introducing the metric in a Lagrangian, thus $\sqrt{-g} = T$ and $g^{tt} = -\frac{1}{T^2}$. With this convention, the parameter $T$ will not appear in the amplitude, 
\ben
A(\phi_f,\phi_i) &=& \<\phi_f\!e^{-2\pi i H}\|\phi_i\>\label{ampli}
\een
Instead $T$ will be implicit in the new way of defining the Hamiltonian, which now will be given by 
\bea
H &=& \frac{T}{2}\int dx \(\pi^2 + \phi^2\)
\eea
With this convention, the form of the amplitude will remain (\ref{ampli}) under the Wick rotation $T=-iR$ and instead the Hamiltonian will change into 
\bea
H &=& -\frac{iR}{2}\int dx \(\pi^2 + \phi^2\)
\eea
The amplitude that we compute does not depend on the convention we use for where to put the parameter $T$. However, it will become important that we use this latter convention when we perform time-reduction of the M5 brane since we want to keep track of the parameter $T$ under the reduction process.

The second ingredient that we now need to introduce is time-reduction, by which we mean that we have a time-like circle on which we dimensionally reduce our theory. We like to stress that this procedure is not strange at all if we use the Lagrangian formulation and we first Wick rotate time $t=-it_E$ and then reduce. In that case, the time direction becomes like a space direction and we can do dimensional reduction just as we are used to do it. However, for the M5 brane where we do not have a classical Euclidean formulation, we can not perform the Wick rotation first. To illustrate time-reduction, let us return to our example on $\mb{R}^{1,1}$ but unlike before we now take $x$ as the direction along which we define Hamiltonian evolution \cite{Hull:2014cxa}. If we want to construct the Hamiltonian associated with translation along Euclidean time $x$, then we shall define the conjugate momentum as
\bea
\pi = \frac{\partial \La}{\partial_x \phi} = -T \partial_x \phi
\eea
and then the Hamiltonian is 
\bea
H &=& \int dt \pi \partial_x \phi - L\cr
&=& \int dt \(-\frac{\pi^2}{2T} - \frac{1}{2T} (\partial_t\phi)^2\)
\eea
The fact that this Hamiltonian is negative is not strange since it is not related to physical energy of the system as $x$ is not a physical time direction. The important point is that the Hamiltonian is bounded (from above in this case). We can see that the physical amplitude (\ref{ampli}) becomes exponentially damped after we take $T=-iR$. The Noether charge that generates translation along $t$ is given by
\bea
P &=& \int dt \pi \partial_t\phi
\eea
We now compactify $t \sim t + 2\pi$. We note that while $H$ is bounded from above, $P$ is unbounded which means that we will find Kaluza-Klein modes of positive and negative mode numbers, just as in the usual situation. Time reduction amounts to keep only the mode number that is equal to zero, or in other words, to put $\partial_t\phi = 0$. We then obtain the time reduced Hamiltonian as
\bea
H &=& - \int_0^{2\pi} dt \frac{1}{2T} \pi^2
\eea
By performing an inverse Legendre transformation, we obtain
\bea
L = \int_0^{2\pi} dt \pi \partial_x \phi - H =  -\int_0^{2\pi} dt \frac{T}{2}(\partial_x\phi)^2
\eea
As expected, this is what we would get if we put $\partial_t \phi = 0$ in the Lagrangian we started with. 

Under the dimensional reduction we also like to rescale the scalar field as
\bea
\Phi &=& 2\pi |T| \phi
\eea
This rescaling is rather harmless since the absolute value $|T|$ means that this rescaling does not imply a further Wick rotation of the scalar field $\Phi$ when $T$ is Wick rotated. Then the time reduced 1d action will be given by
\ben
S &=& -\frac{T}{2\pi|T|^2} \int dx \frac{1}{2}(\partial_x \Phi)^2\label{oneD}
\een
The partition function is computed as
\bea
Z &=& \int \D\Phi e^{i S}
\eea
Wick rotation $T=-iR$ gives 
\ben
Z &=& \int \D\Phi e^{-S_E}\label{1d}
\een
with the Euclidean action
\ben
S_E &=& \frac{1}{2\pi R} \int dt \frac{1}{2} (\partial_x \Phi)^2\label{oneDE}
\een
In this process, we did not need to obtain the Wick rotated 2d Lagrangian. We could derive this Euclidean path integral for the time reduced theory by Wick rotating $T$ in the 1d theory. 

But of course, we can in this example, Wick rotate the 2d theory and start with the Euclidean 2d Lagrangian
\bea
L_E &=& \int dt \frac{R}{2} \(\frac{1}{R^2}(\partial_t\phi)^2 + (\partial_x\phi)^2\)
\eea
Now dimensional reduction along $t$ is nothing strange at all since $t$ is a spatial direction. The Euclidean 2d path integral is defined as
\bea
Z_E &=& \int \D\phi e^{-S_E}
\eea 
with the action $S_E = \int dx L_E$. We dimensionally reduce this action by putting $\partial_t\phi = 0$ and by defining
\bea
\Phi &=& 2\pi R \phi
\eea
We then again obtain the 1d partition function as in (\ref{1d}).

However, inserting the absolute value to prevent $T$ from being Wick rotated seems unnatural. In the Lorentzian theory, $T$ is real (and also positive) so there is no need to insert $|T|$ there. Wick rotation should be carried out at the level of amplitudes, rather than at the level of classical Lagrangians. If $A_L(T)$ denotes the amplitude in Lorentzian theory with $T$ real, then to obtain the corresponding amplitude in Euclidean theory we first analytically continue $T$ from the real axis to the complex plane. Then we rotate $T$ to the imaginary axis and define the Euclidean amplitude as 
\bea
A_E(R) &=& A_L(-iR)
\eea
with $R$ real. In this procedure we never really encounter $|T|$. But computing the amplitude by a path integral requires a choice of integration cycle. We would not like to refer to the choice of integration cycle as Wick rotation. The choice of integration cycle is something more general and there are situations where the choice of integration cycle is not related to Wick rotation of time \cite{Witten:2010zr}. What the insertion of $|T|$ does for us, is to automatically find the new integration cycle after the Wick rotation. But finding the integration cycle is something we always have to do anyway, so we do not really need to invent a device like insertion of various factors of $|T|$ at various places, in order to obtain the new integration cycle. In our application to SYM on five-sphere, we will see that the method of inserting factors of $|T|$ also does not give us the right answer. That the method can fail at some occasions should not come as any surprise. If we like to have a holomorphic partition function $Z(T)$, we can not accept to have some non-holomorphic dependence on $|T| = (T\bar T)^{1/2}$. However, our preliminary presentation becomes more clear by keeping $|T|$, but in the end we will replace $|T|$ with $T$ when $T$ is real, and perform analytical continuation.

Let us go back to our toy example, and see what happens if we write $T$ in place of $|T|$ there. Instead of (\ref{oneD}) we now find
\ben
S &=& -\frac{1}{2\pi T} \int dx \frac{1}{2}(\partial_x \Phi)^2\label{oneDprime}
\een
We use this action to compute the partition function in Lorentzian theory. For some choice of integration cycle of the variable $\Phi$ we obtain a convergent integral that gives us the amplitude $A_L(T)$. In that amplitude we analytically continue $T$ to the complex plane. Let us try to achieve this by analytically continue $T$ to the complex plane directly in the action and finally take $T=-iR$. If we do that, then we end up with the partition function
\bea
Z &=& \int \D \Phi e^{-S_E}
\eea
where 
\ben
S_E &=& -\frac{1}{2\pi R} \int dt \frac{1}{2} (\partial_x \Phi)^2\label{prel}
\een
which for real $\Phi$ is negative definite since we now get the opposite sign compared to what we got in (\ref{oneDE}). But this simply means that we shall choose a different integration cycle of $\Phi$ along the imaginary axis instead of along the real axis. We may write this as that we make the replacement $\Phi \rightarrow i \Phi$ in the above action, and then the new $\Phi$ will be integrated along the real axis. So we end up getting the same answer as we got before.

%\subsection{The M5 brane story}
Our goal is to find the 5d SYM theory that corresponds to Euclidean M5 brane on $S^1 \times S^5$. To this end we start with Lorentzian M5 brane on $\mb{R}\times S^5$ with time along $\mb{R}$. The usual Hamiltonian generates time evolution along $\mb{R}$. But we can define another Hamiltonian as the generator of translation along the fiber direction over $\mb{CP}^2$ inside $S^5$. Declaring this fiber direction as Euclidean time, we can circle compactify $\mb{R}$ into $S^1$ and perform a dimensional reduction along this timelike circle down to 5d. We would then first obtain the Hamiltonian as a density integrated over $\mb{CP}^2$, and then if we perform a Legendre transform, we would get 5d SYM Lagrangian on $S^5$ which will depend on the parameter $T$. In this paper we will take a short-cut that the example we presented above showed us should be possible. Instead of taking the detour via the Hamiltonian, we will obtain the time-reduced Lagrangian directly from 6d by putting time derivatives to zero.

The Wick rotation $T=-iR$ should then take us to the 5d SYM theory that corresponds to Euclidean M5 brane on $S^1\times S^5$ where the radii of these two circles are $R$ and $r$ respectively.

\section{Abelian M5 brane theory}
\subsection{Signatures}\label{signature}
Let us first consider flat M5 brane with global symmetry group $G=$(Lorentz group) $\times $(R-symmetry group). We use the convention that $SO(p,q)$ is the rotation group in $p$ time directions and $q$ space directions. When we write $SO(p)$ instead of $SO(0,p)$. For the arguments we make, we need to introduce our 11d gamma matrices. We will denote these as $\Gamma^M$ and $\hat \Gamma^A$ for the Lorentz group and the R-symmetry group respectively. (We use the index ranges $M=0,1,...,5$ and $A=1,...,5$ in any signatures). They anti-commute $\{\Gamma^M,\hat\Gamma^A\} = 0$. We have the following properties for the 11d charge conjugation matrix,
\bea
(\Gamma^M)^T &=& -C\Gamma^M C^{-1}\cr
(\hat\Gamma^A)^T &=& -C\hat\Gamma^A C^{-1}\cr
C^T &=& -C
\eea
For the Lorentzian M5 brane we will choose $C = \Gamma^0$ which gives the Majorana representation where in addition we find the properties $C^* = C$ and $C^{-1} = -C$.

\subsubsection*{Euclidean 6d theories}
We take $G=SO(6) \times SO(p,5-p)\subset SO(p,11-p)$ for $p=0,1,2,3,4,5$. Let us first examine the 6d Weyl projection. In Eucldean signature we define $\Gamma = i\Gamma^{012345}$. We find that 
\bea
(\Gamma\psi)^{\dag} \hat\Gamma^{1...p} = \psi^{\dag} \Gamma \hat\Gamma^{1...p}=\psi^{\dag}\hat\Gamma^{1...p}\Gamma\cr
(\Gamma\psi)^T C = \psi^T (-C\Gamma C^{-1}) C = - \psi^T C \Gamma
\eea
The minus sign means that a hypothetical 11d Majorana condition $\psi^{\dag} \hat\Gamma^{1...p} = \psi^T C$ can never be imposed on the Weyl components separately. For certain values on $p$ we can also not impose the 11d Majorana condition itself. But we do always have the option of imposing the 6d Weyl condition, if we do not impose or have a Majorana condition. While having complex spinors can be fine, it does not directly relate to 5d theories if we in 5d have real spinors. Therefore we will not consider 6d Euclidean Lagrangians in this paper.

\subsubsection*{Lorentzian 6d theories}
We take $G=SO(1,5)\times SO(p,5-p)\subset SO(1+p,10-p)$. Let us again examine the 6d Weyl projection. In Lorentzian signature we define $\Gamma = \Gamma^{012345}$. We find that
\bea
(\Gamma\psi)^{\dag} \Gamma^0 \hat\Gamma^{1..p} = \psi^{\dag} \Gamma \Gamma^0 \hat\Gamma^{1...p} = -\psi^{\dag}\Gamma^0\hat\Gamma^{1...p} \Gamma\cr
(\Gamma\psi)^T C = -\psi^T C\Gamma
\eea
and so the 6d Weyl projection can always be imposed once we have assured that we have an 11d Majorana spinor, which is the case for some certain values of $p$. 

Let us now examine the possible values on $p$. With $C=\Gamma^0$ we find that $\psi^* = B \psi$ where
\bea
B &=& \hat\Gamma^{1...p}\cr
B^2 &=& (-1)^p (-1)^{\frac{p(p-1)}{2}}
\eea
The first factor $(-1)^p$ comes from that $(\Gamma^{A})^2 = -1$ for each $A=1,...,p$, and the second factor comes from writing $\Gamma^{1...p} = (-1)^{\frac{p(p-1)}{2}} \Gamma^{p...1}$. Then 
\bea
B^* &=& (-1)^p B
\eea
and 
\bea
B^* B &=& (-1)^{\frac{p(p-1)}{2}} 
\eea
Consistency with 11d Majorana condition requires $B^* B = 1$ and therefore
\bea
p(p-1) &\in & 4\mb{Z}
\eea
Solutions are $p=0,1,4,5$ and corresponding R-symmetry groups are $SO(5)$, $SO(1,4)$, $SO(4,1)$ and $SO(5,0)$ which we may imagine as coming from a breaking by the Lorentzian M5 brane of the following 11d Lorentz groups, $SO(1,10)$, $SO(2,9)$, $SO(5,6)$ and $SO(6,5)$ respectively. Essentially all these solutions can be deduced also from the Table $6$ in \cite{Hull:1999mt}. But the case $SO(5,0)$ is exceptional. This table tells us that we can only generate M5 brane with $G=SO(5,1)\times SO(5)$ from M-theory/string theory dualities and reduction processes, but not $G=SO(1,5) \times SO(5,0)$. This can be traced to the possible signatures of M-theory, which are $1+10$, $2+9$ and $5+6$ time plus space dimensions. These are all derived from $1+10$ dimensions by various duality maps. We can not generate $6+5$ dimensional M-theory by these duality maps.

\subsection{Classical field theory description of Lorentzian M5 brane}
Let us assume that we have a smooth Lorentzian six-manifold with metric tensor $g_{MN}$ and signature (-,+,+,+,+,+), which admits some solution to the conformal Killing spinor equation \cite{Linander:2011jy}
\ben
D_M \epsilon &=& \frac{1}{6} \Gamma_M \Gamma^N D_N \epsilon\label{CKS}
\een
Then, by defining $\bar\psi = \psi^{\dag} \Gamma^0$, the following action
\ben
S &=& \int d^6 x \sqrt{-g} {{\cal{L}}},\cr
{{\cal{L}}} &=& \frac{1}{2\pi}\(-\frac{1}{24}H_{MNP} H^{MNP} - \frac{1}{2}\partial_M \phi^A \partial^M \phi_A - \frac{\R}{10}\phi^A \phi_A + \frac{i}{2} \bar\psi\Gamma^M D_M \psi\)\label{6dlagrangian}
\een
is invariant under the following superconformal transformations
\ben
\delta B_{MN} &=& i\bar\epsilon\Gamma_{MN}\psi\cr
\delta \phi^A &=& i\bar\epsilon\Gamma^A\psi\cr
\delta \psi &=& \frac{1}{12}\Gamma^{MNP}\epsilon H_{MNP} + \Gamma^M\Gamma_A\epsilon \partial_M\phi^A - \frac{2}{3}\Gamma_A \Gamma^M D_M \epsilon \phi^A\label{6dvariations}
\een
We lower the R-symmetry indices by the $SO(p,5-p)$ invariant metric 
\bea
\eta_{AB} &=&{\mbox{diag}}(\underbrace{-1,\cdots,-1}_p\underbrace{+1,\cdots,+1}_{5-p})
\eea
For example, if the R-symmetry group is $SO(5,0)$ then we have the 'wrong' sign of the kinetic term
\bea
- \frac{1}{2}\partial_M \phi^A \partial^M \phi_A &=& +\frac{1}{2}\partial_M \phi^A \partial^M \phi^A
\eea
We define the curvature by the relations
\bea
[D_M,D_N]\psi &=& \frac{1}{4}R_{MNAB}\hat\Gamma^{AB}\psi\cr
R_{MN} &=& R_{MPN}{}^P\cr
\R &=& R_M{}^M
\eea
From this and the gamma matrix identity 
\bea
\Gamma^{MN}\Gamma^{AB} &=& -2 g^{MN,AB} - 4 \Gamma^{M[A} g^{B]N} + \Gamma^{MNAB}\cr
\eea
we derive the identity
\bea
\Gamma^{MN} D_M D_N \psi &=& -\frac{1}{4} \R \psi
\eea
which is useful in showing the superconformal invariance. The conserved supercurrent is given by 
\bea
j^M &=& -\frac{i}{12} \epsilon^T C \Gamma^{RST} \Gamma^M \psi H_{RST} - i \epsilon^T C \hat\Gamma_A \Gamma^N \Gamma^M \psi \partial_N \phi^A - 4i D^M \epsilon^T C \hat\Gamma_A \psi \phi^A
\eea
which can be used to derive the superconformal algebra, the stress tensor and central charges.

So far we have not imposed any Weyl condition on the spinor, which is not necessary to show the superconformal invariance of the above action. But for the application to the M5 brane, we need to impose Weyl conditions 
\bea
\Gamma \psi &=& \psi\cr
\Gamma \epsilon &=& -\epsilon
\eea
where $\omega^{MNPQRS} \Gamma = \Gamma^{MNPQRS}$ and $\sqrt{-g}\omega^{012345} = 1$. This leads to the $(2,0)$ tensor multiplet with a selfdual tensor field,
\bea
\frac{1}{6}\omega_{MNP}{}^{RST} H_{RST} &=& H_{MNP}
\eea
Since the tensor field is selfdual, there are some difficulties to write down its corresponding action. One approach could be to let the superconformal current define the theory, and then one may dimensionally reduce this to 5d. Another approach is to work with an action that does not have the full covariance manifest.

\section{A preliminary computation}
Let us now consider the time reduction of $(2,0)$ theory on $\mb{R} \times \mb{R}^5$ with metric 
\bea
ds^2 &=& -(dx^0)^2 + dx^m dx^m
\eea
Let us define the 11d gamma matrices as
\bea
\Gamma^0 &=& i \otimes \sigma^2 \otimes 1\cr
\Gamma^m &=& \gamma^m \otimes \sigma^1\otimes 1\cr
\hat\Gamma^A &=& 1\otimes\sigma^3\otimes\hat\gamma^A
\eea
We then get
\bea
\Gamma := \Gamma^{012345} = 1 \otimes \sigma^3 \otimes 1
\eea
We assume that $\gamma^{12345} = 1 = \hat\gamma^{12345}$. We denote our spinors as $\psi^{\alpha I \dot\alpha}$ where $I=\pm$ for $\Gamma \psi^{\pm} = \pm \psi^{\pm}$ chiralities respectively. We will suppress the index $I$ and write the positive chirality spinor as $\psi^{\alpha \dot\alpha}$ and likewise for the negative chirality supersymmetry parameter we sometimes write this as $\epsilon^{\alpha\dot\alpha}$. 

We define
\bea
\sigma^1 = \(\begin{array}{cc}
0 & 1\\
1 & 0
\end{array}\),\qquad
\sigma^2 = \(\begin{array}{cc}
0 & -i\\
i & 0
\end{array}\),\qquad
\sigma^3 = \(\begin{array}{cc}
1 & 0\\
0 & -1
\end{array}\)
\eea
where the index structures of these matrices is like $(\sigma^1)^I{}_J$ where the upper left corner corresponds to $I=J=+$. We define $\epsilon_{+-} = 1$ and antisymmetric. For the sigma matrices with both indices down, defined as $(\epsilon \sigma^a)_{IJ}= \epsilon_{IK} (\sigma^a)^K{}_J$ we find
\bea
\epsilon\sigma^1 = \(\begin{array}{cc}
1 & 0\\
0 & -1
\end{array}\),\qquad \epsilon\sigma^2 = \(\begin{array}{cc}
i & 0\\
0 & i
\end{array}\),\qquad \epsilon\sigma^3 = \(\begin{array}{cc}
0 & -1\\
-1 & 0
\end{array}\)
\eea
The 11d charge conjugation matrix is 
\bea
C_{11d} &=& C_{\alpha\beta} \epsilon_{IJ} C_{\dot\alpha\dot\beta}
\eea
where $C_{\alpha\beta}$ and $C_{\dot\alpha\dot\beta}$ are antisymmetric. 
 
If we define 5d fields (left-hand sides) in terms of 6d fields (right-hand sides) as
\bea
A_m &=& 2\pi R B_{m0}\cr
\phi^A &=& 2\pi R \phi^A
\eea
and define $F_{mn} = \partial_m A_n - \partial_n A_m$ and we put time derivatives to zero, then the 6d supersymmetry variations reduce to the following 5d supersymmetry variations,
\bea
\delta \phi^A &=& -i \epsilon^{\alpha\dot\alpha} C_{\alpha\beta} (C\gamma^A)_{\dot\alpha\dot\beta} \psi^{\beta\dot\beta}\cr
\delta A_m &=& -i\epsilon^{\alpha\dot\alpha}(C\gamma_m)_{\alpha\beta}C_{\dot\alpha\dot\beta}\psi^{\beta\dot\beta}\cr
\delta \psi^{\alpha\dot\alpha} &=& \frac{1}{2}(\gamma^{mn})^{\alpha}{}_{\beta} \epsilon^{\beta\dot\alpha} F_{mn} - (\gamma^m)^{\alpha}{}_{\beta} (\hat\gamma_A)^{\dot\alpha}{}_{\dot\beta} \epsilon^{\beta\dot\beta} \partial_m \phi^A
\eea
and we find the following supersymmetric the 5d Lagrangian
\bea
{{\cal{L}}}_{5d} &=& \frac{1}{4\pi^2 R} \(\frac{1}{4} F^{mn} F_{mn} - \frac{1}{2} \partial^m \phi^A \partial_m \phi_A + \frac{i}{2} \psi^{\alpha\dot\alpha} (C\gamma^m)_{\alpha\beta} C_{\dot\alpha\dot\beta} \partial_m \psi^{\beta\dot\beta}\)
\eea
The two last terms can be derived by time reduction of the corresponding terms in the 6d Lagrangian
\bea
{{\cal{L}}}_{6d} &=& {{\cal{L}}}_B + \frac{1}{2\pi} \(-\frac{1}{2} \partial^M \phi^A \partial_M \phi_A + \frac{i}{2} \psi^T C \Gamma^M \partial_M \psi\)
\eea
where ${{\cal{L}}}_B$ denotes the Lagrangian of the selfdual tensor field. It is a bit more involved to directly derive the first term in 5d SYM Lagrangian by time reducing the 6d Lagrangian for the selfdual tensor field. It is usually said that no Lagrangian for a selfdual tensor field exists. But if we keep only a subgroup of the Lorentz symmetry manifest, such as $SO(1,2) \times SO(3)$, then we do have a Lagrangian for the selfdual tensor field. In Appendix \ref{A} we present this Lagrangian, together with a derivation of its the time reduction to 5d Euclidean Maxwell theory. After the reduction, the full $SO(5)$ Lorentz symmetry becomes manifest in the reduced Lagrangian.

By looking at the Lagrangian we got above, we can see that if we start with symmetry group  $G=SO(1,5) \times SO(p,5-p)$ in 6d, then upon time reduction we end up with an $SO(5) \times SO(5-p,p)\subset SO(5-p,5+p)$ covariant Lagrangian in 5d. The standard case is $p=0$. This case was considered in \cite{Hull:2014cxa}. The corresponding Euclidean 5d SYM theory one gets by time reduction can be derived from $SO(5,5)$ covariant $\N=1$ SYM in 5+5 dimensions by time reductions along all of its five time directions.

\subsection{Wick rotation}
We define Wick rotation in 5d by taking $A_m = i A^E_m$ while the other fields are not changed. This Wick rotation follows from 6d definition $A_m = 2\pi B_{m0}$ where $B_{m0} = i B_{m0}^E$ and $x^0 = - i x^{E,0}$. Then we get
\bea
\delta \phi^A &=& -i \epsilon^{\alpha\dot\alpha} C_{\alpha\beta} (C\gamma^A)_{\dot\alpha\dot\beta} \psi^{\beta\dot\beta}\cr
\delta A^E_m &=& -\epsilon^{\alpha\dot\alpha}(C\gamma_m)_{\alpha\beta}C_{\dot\alpha\dot\beta}\psi^{\beta\dot\beta}\cr
\delta \psi^{\alpha\dot\alpha} &=& \frac{i}{2}(\gamma^{mn})^{\alpha}{}_{\beta} \epsilon^{\beta\dot\alpha} F^E_{mn} - (\gamma^m)^{\alpha}{}_{\beta} (\hat\gamma_A)^{\dot\alpha}{}_{\dot\beta} \epsilon^{\beta\dot\beta} \partial_m \phi^A
\eea
These leave invariant the Lagrangian
\ben
{{\cal{L}}}_{5d}^E &=& \frac{1}{4\pi^2 R} \(\frac{1}{4} F^{E,mn} F^E_{mn} + \frac{1}{2} \partial^m \phi^A \partial_m \phi_A - \frac{i}{2} \psi^{\alpha\dot\alpha} (C\gamma^m)_{\alpha\beta} C_{\dot\alpha\dot\beta} \partial_m \psi^{\beta\dot\beta}\)
\label{timewick}
\een
We should recall that before the Wick rotation, we had the overall factor of $2\pi$ coming from the integral $\int_0^{2\pi} dx^0$. However this factor was canceled by the factor $1/(2\pi)$ in front of the 6d Lagrangian. But by the Wick rotation, we shall put $x^0=-ix^0_E$. This introduces the extra overall factor of $-i$. It is conventional to define the Euclidean Lagrangian ${{\cal{L}}}^E_{5d} := - i{{\cal{L}}}_{5d}$ whose bosonic part shall be positive definite for the chosen integration cycle.

\section{Time reduction of fields} 
We will now put $T$ in the 6d metric as 
\ben
ds^2 &=& -T^2 dt^2 + G_{mn} dx^m dx^n\label{6dmetric}
\een
where $t\sim t+2\pi$ and $G_{mn}$ is the metric of $S^5$ of radius $r$. 

Let us first recall how we dimensionally reduce a two-form gauge potential along a spatial circle direction $S^1$ characterized by $x^5 \sim x^5 + 2\pi$. We do this by defining a vector potential as
\ben
A^{5d}_{\mu} &=& \int_{S^1} B^{6d}_{\mu 5} dx^5\label{normal}
\een
Then we find that for a closed contour $C$ in the 5d manifold,
\bea
\int_{C} A_{\mu} dx^{\mu} &=& \int_{C \times S^1} B_{\mu 5} dx^{\mu} \wedge dx^5
\eea
The left-hand side is a Wilson line in the 5d theory, the right-hand side is a Wilson surface wrapping the circle on which we dimensionally reduce. Both quantities are well-defined modulo $2\pi$ which fixes the relative normalization in the relation (\ref{normal}). Since in the dimensional reduction we keep only the constant mode, the above definition amounts to\footnote{In the sequel we will omit writing out 'constant mode along $S^1$' in relations like this.}
\bea
A^{5d}_{\mu} &=& 2\pi B^{6d}_{\mu 5}\big|_{\mbox{constant mode along $S^1$}}
\eea
We declare that the same rule holds for dimensional reduction along time,
\bea
A^{5d}_m &=& 2\pi B^{6d}_{m t}\big|_{\mbox{constant mode along time}}
\eea
independently of the 6d metric and in particular independent of the parameter $T$ in the metric (\ref{6dmetric}). 

If we in the Lorentzian theory have a real field $\Phi^{6d}$ which does not carry a vector index along $t$, then we evidently should require the time reduced field to also be real both before and after Wick rotation of $T$. If we also like to have the same scaling with $T$ for both the gauge field and the field $\Phi$, then we shall define 
\bea
\Phi^{5d} &=& 2\pi |T| \Phi^{6d}
\eea
The origin of the scaling by a factor $2\pi |T|$ here will become clear when we come to eq (\ref{Maxwell}) where we see that all terms scale the same way with $T$. Another way to argue for this scaling is that while we can integrate a two-form over the time in a natural way and get a one-form 
\bea
A_m^{5d} &=& \int_0^{2\pi} B_{m t} dt  
\eea
there is no such a natural way that we can integrate a zero form over time, unless we introduce the metric and define
\bea
\sqrt{G} \Phi^{5d} &=& \int_0^{2\pi} dt \sqrt{|g|} \Phi^{6d}
\eea
Since we are going to Wick rotate $T = -i R$, it is essential that we use the scaling factor $2\pi|T|$ and not $2\pi T$ since otherwise $\Phi$ would be Wick rotated which it should not since $\Phi$ did not come from a field that carried any index in the time direction.

\subsection{An equivalent Wick rotation} 
By introducing the parameter $T$ in the 6d metric, a new option arises for how to define Wick rotation in the 5d theory. Instead of tracing how the 5d fields shall be Wick rotated by looking at their 6d origin, we instead only Wick rotate $T = -i R$ in the resulting 5d theory, keeping all the fields fixed. We illustrate the idea only schematically here, where we only keep the components $H_{tmn}$ of the 6d tensor field. For a complete treatment of the selfdual tensor field we refer to the Appendix \ref{A}. Let us define time reduced fields as 
\bea
F^{5d}_{mn} &=& 2\pi H_{tmn}\cr
\phi^{5d} &=& 2\pi |T| \phi
\eea
In our schematic treatment, we define the 6d partition function as
\ben
Z &=& \int \exp i\int_0^{2\pi} dt \int d^5 x \sqrt{-g} \(g^{tt} \(\frac{1}{4} H_{tmn} H_t{}^{mn} + \frac{1}{2} \partial_t \phi \partial_t \phi\) - \frac{1}{2}G^{mn} \partial_m \phi \partial_n \phi\) \cr
\label{6d}
\een
It is important to note that we use $\sqrt{-g}$ in the measure rather than $\sqrt{|g|}$.\footnote{In Lorentzian signature assume that we have some partition function $Z = \int \D\phi \exp i\int dt d^5 x \La(\partial_t \phi)$ for some Lagrangian. Then by standard Wick rotation by taking $t = -it^E$ we get $Z = \int \D\phi \exp \int dt^E d^5 x \La(i\partial_{t^E}\phi)$. If $\La(\partial_t \phi) = \frac{1}{2}(\partial_t\phi)^2$, then $\La(i\partial_{t^E}\phi) = - \frac{1}{2}(\partial_{t^E}\phi)^2$ and $Z = \int \D\phi \exp -\int dt^E d^5 x \frac{1}{2}(\partial_{t^E}\phi)^2$. It is conventional to define the Euclidean Lagrangian so that its kinetic term is positive, as $\La^E = -\La(i\partial_{t^E}\phi)$. Let us now repeat these steps but now instead of Wick rotating $t$, we Wick rotate $T$ in the 6d metric $ds^2 = -T^2 dt^2 + G_{mn} dx^m dx^n$ which has corresponding 6d metric tensor $g_{tt} = -T^2$ and $g_{mn} =G_{mn}$ which has the square root of its determinant as $\sqrt{-g} = T \sqrt{G}$. We start with the 6d partition function $Z = \int \D\phi \exp i \int dt d^5 x \sqrt{-g} \La(\partial_t \phi,T) = \int \D\phi \exp i T\int dt d^5 x \sqrt{G} \La(\partial_t \phi,T)$. Wick rotation by taking $T=-iR$ gives $Z = \int \D\phi \exp R \int dt d^5 x \sqrt{G} \La(\partial_t \phi,-iR)$. With the Lagrangian $\La(\partial_t \phi,T) = -\frac{1}{2}g^{tt} \partial_t \phi \partial_t \phi = \frac{1}{2T^2} \partial_t \phi \partial_t \phi$, Wick rotation gives $\La(\partial_t \phi,-iR) = -\frac{1}{2R^2} \partial_t \phi \partial_t \phi$ and partition function $Z =  \int \D\phi \exp -\frac{1}{2R} \int dt d^5 x \sqrt{G} \partial_t \phi \partial_t \phi$. We thus see that we get the same results as with the other method where we Wick rotate $t = - it^E$. This will not be the case if we choose the measure as $\sqrt{|g|}$ in place of $\sqrt{-g}$.}

Let us now time reduce (\ref{6d}). We get
\ben
Z &=& \int \exp \frac{i}{4\pi^2}\int d^5 x \sqrt{G} \(\frac{1}{4T}  F^{5d}_{mn} F^{5d,mn} - \frac{T}{2|T|^2} \partial_m \phi^{5d} \partial^m \phi^{5d}\)\label{Maxwell}
\een
If we now Wick rotate by taking $T = -iR$, then we get
\bea
Z &=& \int \exp -\frac{1}{4\pi^2 R}\int d^5 x \sqrt{G} \(\frac{1}{4} F^{5d}_{mn} F^{5d,mn} + \frac{1}{2} \partial_m \phi^{5d} \partial^m \phi^{5d}\)
\eea
We see that we ended up with the same result as before in (\ref{timewick}), where we Wick rotated $x^0 = -ix^0_E$.

\section{Lorentzian M5 brane on $S^1\times S^5$}
We take the Lorentzian 6d metric on $\mb{R} \times S^5$ as  
\bea
ds^2 &=& -T^2 dt^2 + G_{mn} dx^m dx^n
\eea
where we introduce a parameter $T$ and where $G_{mn}$ denotes the metric of the $S^5$ of radius $r$. The most general solution to (\ref{CKS}) is given by
\bea
\epsilon &=& e^{i\frac{T}{2r}t} \E(x^m) + e^{-i\frac{T}{2r}t} \F(x^m)
\eea
where
\bea
D_m \E &=& i\frac{T}{2r} \Gamma_m \Gamma^t \E\cr
D_m \F &=& -i\frac{T}{2r} \Gamma_m \Gamma^t \F
\eea
We define 
\bea
\Gamma^t &=& \frac{1}{T} \Gamma^0
\eea
where $(\Gamma^0)^2 = -1$, as defined in the previous section. 

Let us define a group element
\bea
g &=& \exp i \(\Lambda^{12} \hat M_{12} + \Lambda^{34} \hat M_{34}\)
\eea
Here $\hat M_{12}$ and $\hat M_{34}$ are Cartan generators of $SO(2)\times SO(2)$ inside $SO(5)$ R-symmetry group. In the spinor representation $\hat M_{34} = \frac{i}{2} \hat\Gamma_{34}$ and in the vector representation $(\hat M_{34})^{AB} = 2 i \delta^{AB}_{34}$.

Let us use a spin basis and define
\bea
-\frac{i}{2} \hat\Gamma_{12} \epsilon^{s_1 s_2} &=& s_1 \epsilon^{s_1 s_2}\cr
-\frac{i}{2} \hat\Gamma_{34} \epsilon^{s_1 s_2} &=& s_2 \epsilon^{s_1 s_2}
\eea
Then 
\bea
g \epsilon^{s_1 s_2} &=& e^{-i \(\Lambda^{12} s_1 + \Lambda^{34} s_2\)} \epsilon^{s_1 s_2}\cr
&=& e^{i \(-\Lambda^{12} s_1 - \Lambda^{34} s_2 + \frac{T}{2r} t\)} \E^{s_1 s_2} + e^{i\(-\Lambda^{12} s_1 - \Lambda^{34} s_2 - \frac{T}{2r}t\)} \F^{s_1 s_2}
\eea
Let us make the ansatz 
\ben
\Lambda^{12} &=& \(\frac{T}{2r} - \lambda\) t\cr
\Lambda^{34} &=& \(\frac{T}{2r} + \lambda\) t\label{param}
\een
as real-valued gauge parameters, after gauging this $SO(2)\times SO(2)$ R-symmetry. Then we get
\bea
g \epsilon^{s_1 s_2} &=& e^{i t \(\frac{T}{2r} (-s_1 - s_2 + 1) + \lambda (-s_2 + s_1)\)} \E^{s_1 s_2} + e^{i t \(\frac{T}{2r} (-s_1 - s_2 - 1) + \lambda (-s_2 + s_1)\)} \F^{s_1 s_2}
\eea
If $\lambda \neq \pm \frac{T}{2r}$, then we find that $t$-dependence is canceled for the components $\E^{++}$ (i.e. $s_1 = s_2 = +\frac{1}{2}$) and $\F^{--}$, which then will be the surviving supersymmetries in 5d. If $\lambda = \frac{T}{2r}$, then we find additional supersymmetries $\E^{-+}$ and $\F^{+-}$, and likewise if $\lambda = -\frac{T}{2r}$ we find instead additional supersymmetries $\E^{+-}$ and $\F^{-+}$. These cases when $\lambda = \pm \frac{T}{2r}$ correspond to cases when we find manifest $\N=2$ supersymmetry (see \cite{Kim:2012ava} for details), while the generic case only gives manifest $\N=1$ supersymmetry.

We define
\bea
\partial_M (g^{-1} \Phi) &=& g^{-1} D_M \Phi
\eea
For an explanation why the inverse element $g^{-1}$ appears here, we refer to Appendix \ref{B}. This way we find that 
\bea
D_t \Phi &=& \partial_t \Phi + g \partial_t g^{-1} \Phi
\eea
Upon the reduction we put $\partial_t = 0$. Then 
\bea
D_t \Phi &=& \(-i \(\frac{T}{2r} - \lambda\) \hat M_{12} - i \(\frac{T}{2r} + \lambda\) \hat M_{34} \)\Phi
\eea
Explicitly, and if $a=1,2$ and $i=3,4$, then
\bea
D_t \psi &=& \(\frac{1}{2} \(\frac{T}{2r} - \lambda\) \hat \Gamma_{12} + \frac{1}{2} \(\frac{T}{2r} + \lambda\) \hat \Gamma_{34} \) \psi\cr
D_t \phi^a &=& \(\frac{T}{2r} - \lambda\) \epsilon^{ab} \phi^b\cr
D_t \phi^i &=& \(\frac{T}{2r} + \lambda\) \epsilon^{ij} \phi^j
\eea
where $\epsilon^{ab} = 2 \delta^{ab}_{12}$ and $\epsilon^{ij} = 2\delta^{ij}_{34}$. 

We now find 5d mass terms from the following terms in the 6d theory,
\bea
-\frac{1}{2} g^{tt} D_t \phi^a D_t \phi^a &=& \frac{1}{2}\(\frac{1}{2r} - \frac{\lambda}{T}\)^2 \phi^a \phi^a\cr
-\frac{1}{2} g^{tt} D_t \phi^i D_t \phi^i &=& \frac{1}{2}\(\frac{1}{2r} + \frac{\lambda}{T}\)^2 \phi^i \phi^i
\eea
in addition to the 6d conformal mass
\bea
-\frac{2}{r^2} \(\phi^a \phi^a + \phi^i \phi^i + \phi^5 \phi^5\)
\eea
Adding up, we have the mass terms
\bea
\(-\frac{15}{8r^2} + \frac{\lambda^2}{2T^2}\) \(\phi^a \phi^a + \phi^i \phi^i\) + \frac{\lambda}{2rT} \(\phi^i\phi^i - \phi^a \phi^a\)
\eea
for the hypermultiplet, and 
\bea
-\frac{2}{r^2} \phi^5 \phi^5
\eea
for the vector multiplet.

Let us also expand the fermionic term. To this end we find it convenient to choose the Weyl representation of the $SO(4)$ gamma matrices
\bea
\hat\gamma^A &=& \(\begin{array}{cc}
0 & \sigma^A\\
\bar\sigma^A & 0
\end{array}\)
\eea
where
\bea
\sigma^A &=& (\sigma^1,\sigma^2,\sigma^3,-i)\cr
\bar\sigma^A &=& (\sigma^1,\sigma^2,\sigma^3,i)
\eea
Then 
\bea
\hat\gamma = \hat\gamma^{1234} = \(\begin{array}{cc}
-1 & 0 \\
0 & 1
\end{array}\)
\eea
We also decompose the spinor accordingly into Weyl components
\bea
\psi &=& \(\begin{array}{c}
\psi_+\\
\psi_-
\end{array}\)
\eea
Then 
\bea
\hat\gamma_{12} = i\(\begin{array}{cc}
\sigma^3 & 0\\
0 & \sigma^3
\end{array}\),\qquad \hat\gamma_{34} = i\(\begin{array}{cc}
\sigma^3 & 0\\
0 & -\sigma^3
\end{array}\)
\eea
We then get
\bea
\frac{i}{2} \psi^T C \Gamma^t D_t \psi &=& \frac{1}{4r} \psi^T_+ \epsilon\sigma^3 \psi_+ + \frac{\lambda}{2T} \psi^T_- \epsilon \sigma^3 \psi_-
\eea

The full Lagrangian is a sum of three terms, the kinetic terms, the Scherk-Schwarz mass and the conformal mass terms. Adding them up, and taking into account rescaling of the fields under time reduction as previously discussed, we find the Lagrangian as 
\ben
\La &=& \La^{vector} + \La^{hyper}\label{minkow}
\een
where
\bea
4\pi^2 \La^{vector} &=& \frac{1}{4T} F_{mn} F^{mn} - \frac{T}{2|T|^2} \partial_m \phi^5 \partial^m \phi^5 - \frac{T}{|T|^2} \frac{2}{r^2} \phi^5 \phi^5\cr
&&+ \frac{i}{2} \frac{T}{|T|^2}  \psi_a \epsilon^{ab} \gamma^m D_m \psi_b + \frac{1}{4r} \frac{T}{|T|^2}  \psi_a (\epsilon \sigma^3)^{ab} \psi_b
\eea
and
\bea
4\pi^2 \La^{hyper} &=& -\frac{T}{2|T|^2} \(\partial^m \phi^a \partial_m \phi^a + \partial^m \phi^i \partial_m \phi^i\) \cr
&&+ \(-\frac{15 T}{8r^2 |T|^2} + \frac{T}{2|T|^2} \(\frac{\lambda}{T}\)^2\) \(\phi^a \phi^a + \phi^i \phi^i\)\cr
&&+ \frac{1}{2|T|^2} \frac{\lambda}{r} \(-\phi^a \phi^a + \phi^i\phi^i\)\cr
&&+ \frac{i}{2}\frac{T}{|T|^2}   \psi^r \epsilon_{rs} \gamma^m D_m \psi^r + \frac{\lambda}{2T} \frac{T}{|T|^2}  \psi^r (\epsilon \sigma^3)_{rs} \psi^s
\eea
This 5d SYM Lagrangian corresponds to M5 brane on $\mb{R} \times S^5$ with the usual $SO(5)$ R-symmetry, which has been broken down to $SO(2) \times SO(2)$ by the timelike holonomy. 

Since $T$ is real in the Lorentzian case, the above Lagrangian in Lorentzian signature, reduces to 
\ben
\La &=& \frac{1}{4\pi^2 T} \(\La^{vector} + \La^{hyper}\)\label{minkow1}
\een
where
\bea
\La_{vector} &=& \frac{1}{4} F_{mn} F^{mn} - \frac{1}{2} \partial_m \phi^5 \partial^m \phi^5 - \frac{2}{r^2} \phi^5 \phi^5\cr
&&+ \frac{i}{2} \psi_a \epsilon^{ab} \gamma^m D_m \psi_b + \frac{1}{4r} \psi_a (\epsilon \sigma^3)^{ab} \psi_b
\eea
and
\bea
 \La_{hyper} &=& -\frac{1}{2} \(\partial^m \phi^a \partial_m \phi^a + \partial^m \phi^i \partial_m \phi^i\) \cr
&&+ \(-\frac{15}{8r^2} + \frac{1}{2} \(\frac{\lambda}{T}\)^2\) \(\phi^a \phi^a + \phi^i \phi^i\)\cr
&&+ \frac{1}{2r} \frac{\lambda}{T} \(-\phi^a \phi^a + \phi^i\phi^i\)\cr
&&+ \frac{i}{2}  \psi^r \epsilon_{rs} \gamma^m D_m \psi^r + \frac{\lambda}{2T} \psi^r (\epsilon \sigma^3)_{rs} \psi^s
\eea

At $\lambda = \frac{T}{2r}$ and with $T$ real, we have enhanced $SO(3) \times SO(2)$ R-symmetry, $\N=2$ supersymmetry and the Lagrangian
\ben
4\pi^2 T \La &=& \frac{1}{4} F_{mn} F^{mn} - \frac{1}{2} \( \partial^m \phi^i \partial_m \phi^i +  \partial^m \phi^a \partial_m \phi^a + \partial_m \phi^5 \partial^m \phi^5\) \cr
&&- \frac{3}{2r^2} \phi^i \phi^i - \frac{4}{2r^2} \(\phi^a \phi^a + \phi^5 \phi^5\)\cr
&&+ \frac{i}{2} \psi_a \epsilon^{ab} \gamma^m D_m \psi_b + \frac{i}{2} \psi^r \epsilon_{rs} \gamma^m D_m \psi^r \cr
&& + \frac{1}{4r} \(\psi_a (\epsilon \sigma^3)^{ab} \psi_b + \psi^r (\epsilon \sigma^3)_{rs} \psi^s\)\label{Seok}
\een

\section{Euclidean M5 brane on $S^1 \times S^5$}
Let us Wick rotate $T = -iR$ and let us also make the replacement $\phi^A \rightarrow i \phi^A$ for $A=(i,a,5)= 1,2,3,4,5$ in (\ref{minkow1}). We then get the following Euclidean 5d SYM Lagrangian that correponds to Euclidean M5 brane, 
\ben
\La^E &=& \frac{1}{4\pi^2 R} \(\La^E_{vector} + \La^E_{hyper}\)\label{EUCL}
\een
where
\bea
\La^E_{vector} &=& \frac{1}{4} F_{mn} F^{mn} + \frac{1}{2} \partial_m \phi^5 \partial^m \phi^5 + \frac{2}{r^2} \phi^5 \phi^5\cr
&&- \frac{i}{2} \psi_a \epsilon^{ab} \gamma^m D_m \psi_b - \frac{1}{4r} \psi_a (\epsilon \sigma^3)^{ab} \psi_b
\eea
and
\bea
\La^E_{hyper} &=& \frac{1}{2} \(\partial^m \phi^a \partial_m \phi^a + \partial^m \phi^i \partial_m \phi^i\) \cr
&&+ \(\frac{15}{8r^2} - \frac{1}{2} \(\frac{i\lambda}{R}\)^2\) \(\phi^a \phi^a + \phi^i \phi^i\)\cr
&&+ \frac{1}{2 R} \frac{i\lambda}{r} \(\phi^a \phi^a - \phi^i\phi^i\)\cr
&&- \frac{i}{2} \psi^r \epsilon_{rs} \gamma^m D_m \psi^r - \frac{i\lambda}{2R} \psi^r (\epsilon \sigma^3)_{rs} \psi^s
\eea
We can now identify the hypermultiplet as
\bea
m_{hyper} &=& \frac{i\lambda}{R}
\eea
which is purely imaginary. In the $\N=1$ language there is an argument that says that $m_{hyper}$ shall be Wick rotated along with the vector multiplet scalar field $\phi^5$ \cite{Minahan:2013jwa}, since it is identified as the vacuum expectation value of a vector multiplet scalar field \cite{Hosomichi:2012ek}. Also, in order to have a well-defined localization locus as can be seen from eq (3.25) in \cite{Hosomichi:2012ek}, we have to rotate the hypermultiplet mass at the same time as we rotate $\phi^5$ into the imaginary axis. This also seems to fit well into our picture. In the Lorentzian theory it appears like the scalar field kinetic terms have the wrong sign. But that is an artifact of the time reduction from the Lorentzian M5 brane theory. We should thus consider the scalar fields with wrong sign kinetic terms as real valued fields. By Wick rotating $T$ into Euclidean theory, our preliminary example and eq (\ref{prel}) suggests that we should also Wick rotate the scalar fields into the imaginary axis, thus providing an independent argument why we should Wick rotate the hypermultiplet mass (which is automatically being Wick rotated as we Wick rotate $T$, since $m_{hyper} = \frac{\lambda}{T}$ in the Lorentzian case) simultaneously with the vector multiplet scalar field. 

We can use the Lagrangian (\ref{EUCL}) for the localization computation. It agrees with the Lagrangian that was proposed in \cite{Minahan:2013jwa} for the Euclidean M5 brane, and we show this identification in more detail in the Appendix \ref{C}.\footnote{In \cite{Minahan:2013jwa} the rotation of the hypermultiplet scalars was not considered, and hence they found the other real slice which has $SO(1,2) \times SO(2)$ R symmetry instead of $SO(3)\times SO(2)$.} 

The localization computation can be done in Euclidean theory, but the unitary physical theory is in Lorentzian signature, and that Lagrangian is obtained by Wick rotating $T$ as well as $\phi^5$ and the four hypermultiplet scalars. We then obtain the Lorentzian 5d SYM Lagrangian in eq (\ref{minkow1}), which is real. It has a non-Abelian generalization, which is also real. There will in particular be a cubic interaction term that is real, but which becomes purely imaginary upon Wick rotation of the five scalar fields. This purely imaginary cubic interaction term was first noted in \cite{Kim:2012ava}.

We can not keep manifest $\N=2$ supersymmetry in the Lagrangian under Wick rotation. The parameter $\lambda$ before Wick rotation is real and for enhanced $\N=2$ supersymmetry this shall be taken at either of the two critical values 
\bea
\lambda &=& \pm \frac{T}{2r}
\eea
in which case we get
\bea
m_{hyper} &=& \pm \frac{1}{2r}
\eea
After the Wick rotation $\lambda$ will remain at the same real value. Hence the critical values after the Wick rotation are at
\bea
\lambda &=& \pm \frac{R}{2r}
\eea
in which case we get
\bea
m_{hyper} &=& \pm \frac{i}{2r}
\eea
At these values we can only have $\N=1$ supersymmetry manifest in the Lagrangian. But at the quantum level we do not break any supersymmetry by Wick rotating $T$. If some supercharge commutes with the Hamiltonian before Wick rotation, then this will be true also after Wick rotation of $T$.

\section{Discussion}
If we Wick rotate $T$ in the Lagrangian (\ref{minkow}) then we automatically get (\ref{EUCL}) without rotating the integration contour as $\Phi^A \rightarrow i \Phi^A$. However, there is a problem with this approach. It appears to us that the gauge field $A_t$ should not be Wick rotated as we Wick rotate $T$. The metric is $ds^2 = -T^2 dt^2 + ...$ so if we Wick rotate $t$ as well as Wick rotate $T$, then we are not doing anything sensible. Now if we choose the gauge parameter as in eq (\ref{param}), then as $T$ is Wick rotated to the imaginary axis, also the gauge field $A_t$ will become complexified. But if we would replace $T$ by $|T|$ in (\ref{param}) to make sure this is not being Wick rotated, then Wick rotation of $T$ would break all the manifest supersymmetry, and we would not get the right answer. This kind of problem is expected when we try to understand what happens to a field such as $A_t$ in the classical 6d theory as we Wick rotate to Euclidean signature, because there is no Euclidean 6d theory at the level of a classical Lagrangian. 

The right way to to proceed is by first computing the partition function $Z_L(T,r,\lambda)$ in Lorentzian M5 brane theory using Hamiltonian quantization, thus keeping $T,r,\lambda$ as arbitrary but real parameters. We then analytically continue $T$ to the complex plane, and then the Euclidean partition function will be given by $Z_E(R,r,\lambda) = Z_L(-iR,r,\lambda)$. 

At the critical points $\lambda = \pm \frac{T}{2r}$ we have manifest $\N=2$ supersymmetry in the Lorentzian theory. The partition function is given by $\t Z_L(T,r) = Z_L(T,r,\pm\frac{T}{2r})$ with $T$ and $r$ real numbers. But to obtain the Euclidean partition function, we shall not analytically continue $\t Z_L(T,r)$ in $T$ and then define the Euclidean partition function as $\t Z_E(R,r) = \t Z_L(-iR,r)$. We shall keep $\lambda$ real and continue $T$ analytically, so for $\lambda = \pm \frac{T}{2r}$ we get the Euclidean partition function as $Z_E(R,r) = Z_L(-iR,r,\lambda = \pm \frac{R}{2r})$ where $R$ is real and positive. 

Let us summarize: In the Lorentzian theory, we have a partition function of a real parameter $\lambda$ and a real time interval $T$. For a generic real value of $\lambda$ we have $\N=1$ 5d SYM theory, that is obtained by time reduction of 6d theory with $8$ superconformal charges. To reach the corresponding Euclidean 6d theory, we analytically continue in $T$ only. Hence we keep $\lambda$ fixed at its real value since in general it is not related to $T$, but is an unrelated free parameter. If for some certain value of $\lambda$ we had an enhanced symmetry in the Lorentzian signature (say enhancement to $16$ superconformal charges), then this will remain true also after the Wick rotation, at the same real value of $\lambda$. This will be true even if this is not manifest in the Lagrangian formulation after the Wick rotation of $T$.

\vskip20pt

\subsection*{Acknowledgements}
I would like to thank Dongsu Bak, Kimyeong Lee and Soo-Jong Rey for discussions. This work was supported in part by NRF Grant 2014R1A1A2053737.

\newpage
\appendix
\section{Euclidean quantum field theory}\label{Q}
Here we study the amplitude 
\bea
A(\phi_N,\phi_0) &=& \<\phi_N\!e^{-2\pi R H}\|\phi_0\>
\eea
using the Hamiltonian in (\ref{haminapp}) that we derived from the Lorentzian Lagrangian, with the goal being to derive the corresponding Euclidean Lagrangian. We discretize the interval $2\pi R$ into $N$ segments, each of lenght $\epsilon$ and write $e^{-2\pi RH} = e^{-\epsilon H} e^{-\epsilon H} \cdots e^{-\epsilon H}$. We then insert a complete set of states $1 = \int \|\phi_i\>\<\phi_i\!$ between each factor. Then one such factor associated to segment from $i$ to $i+1$ is given by
\bea
\<\phi_{i+1}\!e^{-\epsilon H}\|\phi_i\> &=& \int d\pi_i \<\phi_{i+1}\!e^{-\epsilon \int \frac{1}{2}\pi^2} \|\pi_i\>\<\pi_i\!e^{-\epsilon\int \frac{1}{2}(\partial_x \phi)^2}\|\phi_i\>\cr
&=& \int d\pi_i e^{i(\phi_{i+1}-\phi_i)\pi_i} e^{-\epsilon \int \frac{1}{2}\(\pi^2 + (\partial_x\phi)^2\)}
\eea
The first exponent is determined by the equal time commutation relation
\bea
[\phi,\pi] &=& i
\eea
Let us now make the ansatz 
\bea
\phi_{i+1} - \phi_i &=& \phi'_i \epsilon
\eea
where the prime denotes derivative with respect to the variable along which $H$ evolves. We then get
\bea
=\int d\pi_i e^{\epsilon \(i\phi'_i \pi_i - \frac{1}{2}\(\pi_i^2 + (\partial_x\phi)^2\)\)} &=& e^{-\int dx \frac{1}{2} \({\phi'}_i^2 + (\partial_x\phi_i)^2\)}
\eea
where we integrated out $\pi_i$ that put 
\bea
\pi_i &=& i \phi'_i
\eea
By multiplying all these segment contributions and integrating over each $\phi_i$, we obtain the path integral of the Euclidean Lagrangian
\bea
L_E &=& \frac{1}{2}\(\phi'^2 + (\partial_x\phi)^2\)
\eea
But we started with the Hamiltonian and the canonical commutation relation that we derived from the Lorentzian Lagrangian $L$. The only thing we did 'wrong', was that we computed 
$\tr\(e^{-R H}\)$ instead of $\tr\(e^{-iTH}\)$. Had we computed the latter quantity instead, we would have got the path integral over the Lorentzian action we started with. And indeed, the prime corresponds to Euclidean time derivative $\phi' = \partial\phi/\partial t_E$ where $t=-it_E$.

\section{Time reduction for selfdual tensor field}\label{A}
On flat $\mb{R}^{1,5}$, there is an $SO(1,2)\times SO(3)$ covariant Lagrangian for a selfdual tensor field. If we let $\mu \in SO(1,2)$ and $\alpha \in SO(3)$ be the vector indices of the reduced 6d Lorentz group, then the 6d Lagrangian can be written as
\bea
{{\cal{L}}} &=& \frac{1}{2\pi}\(-\frac{1}{2} H_{\mu\alpha\beta} H^{-,\mu\alpha\beta} - \frac{1}{6} H_{\alpha\beta\gamma} H^{-,\alpha\beta\gamma}\)
\eea
where we define
\bea
H^{-,\mu\alpha\beta} &=& \frac{1}{2} \(H^{\mu\alpha\beta} - \frac{1}{2} \epsilon^{\mu\nu\lambda} \epsilon^{\alpha\beta\gamma} H_{\nu\lambda\gamma}\)\cr
H^{-,\alpha\beta\gamma} &=& \frac{1}{2}\(H^{\alpha\beta\gamma} - \frac{1}{6} \epsilon^{\mu\nu\lambda} \epsilon^{\alpha\beta\gamma} H_{\mu\nu\lambda}\)
\eea
As argued in \cite{Pasti:2009xc} this action corresponds to the equation of motion $H^- = 0$, which means that this is an action describing the dynamics of a selfdual three-form $H^+$. 

In this Lagrangian, all terms which involve $B_{\mu\nu}$ are total derivatives. If we ignore those total derivatives, then this Lagrangian can also be expressed without making use of $B_{\mu\nu}$ components as
\bea
{{\cal{L}}} &=& \frac{1}{2\pi}\(-\frac{1}{4} H_{\mu\alpha\beta} H^{\mu\alpha\beta} - \frac{1}{12} H_{\alpha\beta\gamma} H^{\alpha\beta\gamma} + \frac{1}{2} \epsilon^{\mu\nu\lambda}\epsilon^{\alpha\beta\gamma} \partial_{\alpha} B_{\beta\mu} \partial_{\nu} B_{\lambda \gamma}\)
\eea
If we use the flat but rescaled metric $ds^2 = -T^2 dt^2 + dx^m dx^m$, then we will replace $\epsilon^{\mu\nu\lambda}$ with the covariant tensor $\omega^{\mu\nu\lambda}$ which we define such that $T \omega^{t12} = 1$. Being covariant means that $\omega_{t12} = g_{tt} \omega^{t}{}_{12} = - T$.

\subsection{The time reduction}
We will now show that time reducing this Lagrangian, we obtain the $SO(5)$ covariant 5d Maxwell Lagrangian. We decompose the $SO(1,2)$ vector index as $\mu = (t,i)$ where $i=1,2$. We put $\partial_t = 0$ and we define 
\bea
F_{mn} &=& 2\pi H_{tmn}
\eea
It is then important to note that $H^{tmn} = g^{tt} F^{mn} = -\frac{1}{2\pi T^2} F^{mn}$. We then get
\bea
2\pi {{\cal{L}}} &=& -\frac{1}{12} H_{\alpha\beta\gamma}H^{\alpha\beta\gamma} + \frac{1}{4}\frac{1}{4\pi^2 T^2} F_{\alpha\beta} F^{\alpha\beta} - \frac{1}{4} H_{i\alpha\beta} H^{i\alpha\beta} \cr
&& + \frac{1}{2} \omega^{tij} \epsilon^{\alpha\beta\gamma} \(\partial_{\alpha} B_{\beta 0} \partial_{i}B_{j\gamma} + \partial_{\alpha} B_{\beta j} \partial_i B_{0\gamma}\)
\eea
We perform integration by parts, and bring the last two terms into the form
\bea
+\omega^{tij} \epsilon^{\alpha\beta\gamma} \partial_{\alpha} B_{\beta j} \partial_i B_{0\gamma} &=& -\frac{1}{2}\omega^{tij} \epsilon^{\alpha\beta\gamma} H_{\alpha\beta j} \partial_i B_{\gamma 0}
\eea
where we used $[\partial_i,\partial_j] = 0$ and subtracted another total derivative term. We expand this term as
\bea
&=& -\frac{1}{2}\omega^{tij} \epsilon^{\alpha\beta\gamma} H_{\alpha\beta j} \(\frac{1}{2\pi} F_{i\gamma} + \partial_{\gamma} B_{i 0}\)
\eea
Let us now look at the second term here,
\bea
-\frac{1}{2}\omega^{tij} \epsilon^{\alpha\beta\gamma} H_{\alpha\beta j} \partial_{\gamma} B_{i 0} &=& \frac{1}{2}\omega^{tij} \epsilon^{\alpha\beta\gamma} \partial_{\gamma} H_{\alpha\beta j} B_{i0}
\eea
We now use the Bianchi identity
\bea
\partial_{[\gamma} H_{\alpha\beta] j} &=& \frac{1}{3}\partial_j H_{\alpha\beta\gamma} \eea
and we get this term as
\bea
\frac{1}{6}\omega^{tij} \epsilon^{\alpha\beta\gamma} \partial_{j} H_{\alpha\beta\gamma} B_{i0} &=& \frac{1}{12}\omega^{tij} \epsilon^{\alpha\beta\gamma} H_{\alpha\beta\gamma} \frac{1}{2\pi} F_{ij}
\eea
Summing up, we have
\bea
-\frac{1}{2}\omega^{tij} \epsilon^{\alpha\beta\gamma} H_{\alpha\beta j} \partial_i B_{\gamma 0} &=& -\frac{1}{2\pi} \omega^{tij} \epsilon^{\alpha\beta\gamma} \(\frac{1}{2}H_{\alpha\beta j} F_{i\gamma} - \frac{1}{12} H_{\alpha\beta\gamma} F_{ij}\)
\eea
and the full Lagrangian is
\bea
2\pi {{\cal{L}}} &=& -\frac{1}{12} H_{\alpha\beta\gamma}H^{\alpha\beta\gamma} + \frac{1}{4} \frac{1}{4\pi^2 T^2} F_{\alpha\beta} F^{\alpha\beta} - \frac{1}{4} H_{i\alpha\beta} H^{i\alpha\beta} \cr
&& - \frac{1}{2\pi} \omega^{tij} \epsilon^{\alpha\beta\gamma} \(\frac{1}{2}H_{\alpha\beta j} F_{i\gamma} - \frac{1}{12} H_{\alpha\beta\gamma} F_{ij}\)
\eea
Let us now consider the $SO(1,5)$ covariant selfduality constraint 
\bea
H_{MNP} &=& \frac{1}{6} \omega_{MNP}{}^{QRS} H_{QRS}
\eea
which implies that 
\bea
H_{\alpha\beta\gamma} &=& \frac{1}{6} \epsilon_{\alpha\beta\gamma} \omega_{\mu\nu\lambda} H^{\mu\nu\lambda}\cr
H_{\mu\alpha\beta} &=& -\frac{1}{2} \epsilon_{\alpha\beta\gamma} \omega_{\mu\nu\lambda} H^{\nu\lambda\gamma}
\eea
The minus sign in the second equation arises as follows
\bea
\omega_{\mu\alpha\beta\nu\lambda\gamma} = -\omega_{\alpha\beta\gamma\mu\nu\lambda} = -\omega_{\alpha\beta\gamma}\omega_{\mu\nu\lambda}
\eea
We will be interested in these relations in the form
\bea
H_{\alpha\beta\gamma} &=& \frac{1}{2} \omega_{\alpha\beta\gamma} \omega_{t ij} H^{t ij}\cr
H_{i \alpha\beta} &=& -\omega_{ij t} \omega_{\alpha\beta\gamma} H^{t\gamma j}
\eea
While we do not intend to impose selfduality like this by hand in 6d, this nevertheless motivates us to make the following field redefinitions in 5d,
\bea
H_{\alpha\beta\gamma} &=& \frac{1}{4\pi T^2} \omega_{\alpha\beta\gamma} \omega_{t ij} F^{ij}\cr
H_{i \alpha\beta} &=& -\frac{1}{2\pi T^2} \omega_{ij t} \omega_{\alpha\beta\gamma} F^{\gamma j}
\eea
The idea here is that a three-form in 5d can be dualized into a two-form. In 5d these relations are not constraining the tensor or vector fields, but are merely field redefinitions which we freely can perform. They should be chosen precisely this way because this choice will give us a manifestly $SO(5)$ covariant 5d Lagrangian. Making the above field redefinitions, we get
\bea
2\pi {{\cal{L}}} &=& \frac{1}{4\pi^2 T^2} \(\frac{1}{4} F_{ij} F^{ij} + \frac{1}{2} F_{i\alpha} F^{i\alpha} + \frac{1}{4} F_{\alpha\beta} F^{\alpha\beta}\)
\eea
To obtain the 5d Lagrangian we need to integrate the 6d Lagrangian over time which produces an overall factor $2\pi T$. We end up with
\bea
{{\cal{L}}}_{5d} &=& \frac{1}{4\pi^2 T} \frac{1}{4} F_{mn} F^{mn}
\eea
which the manifestly $SO(5)$ covariant 5d Maxwell Lagrangian. We have derived the first term in Eq (\ref{Maxwell}) in the main text.

\section{Gauging the R-symmetry}\label{B}
The global R-symmetry can  be gauged by introducing a R-symmetry gauge field and corresponding covariant derivative. This does not affect the supersymmetry at all if we demand the R-symmetry gauge field to be invariant under supersymmetry. The global R-symmetry acts on all fields that are charged under the R-symmetry, as
\ben
\psi &\rightarrow & g^{-1} \psi\cr
\phi^A &\rightarrow & (g^{-1})^A{}_B \phi^B\label{Rfield}
\een
where 
\bea
g &=& \exp \frac{i}{2} \Lambda^{AB} \hat M_{AB}
\eea
is an element of the R-symmetry group with generators $M_{AB}$ and real antisymmetric constant parameters $\Lambda^{AB}$. Let us now illustrate how R-symmetry transformation of the fields can be traded for a transformation of the supersymmetry parameter, by considering the following part of a supersymmetry variation,
\bea
\delta \psi &=& \Gamma^M \Gamma_A \epsilon \partial_M \phi^A
\eea
We now perform the R-symmetry transformation (\ref{Rfield}) on the fields, to get
\bea
\delta \psi &=& g \Gamma^M \Gamma_A g^{-1} (g\epsilon) \partial_M \( (g^{-1})^A{}_B\phi^B\)
\eea
Let us now also promote the parameters $\Lambda^{AB}$ to local functions, and expand out 
\bea
\partial_M \( (g^{-1})^A{}_B\phi^B\) &=&  (g^{-1})^A{}_B  D_M \phi^B
\eea
where we shall define
\bea
D_M \phi^B &=& \partial_M \phi^B - i (A_M)^B{}_D \phi^D\cr
(A_M)^B{}_D &=& i g^B{}_C \partial_M(g^{-1})^C{}_D
\eea
We plug this back in and get
\bea
\delta \psi &=& g \Gamma^M \Gamma_A g^{-1}  (g^{-1})^A{}_B (g\epsilon)  D_M \phi^B
\eea
We indeed have the invariance condition
\bea
g \Gamma_A g^{-1} (g^{-1})^A{}_B &=& \Gamma_B
\eea
so we find that 
\bea
\delta \psi &=& \Gamma^M \Gamma_B (g\epsilon)  D_M \phi^B
\eea
which shows that the R-symmetry transformation (\ref{Rfield}) as far as supersymmetry variations concern, equivalently can be obtained by transforming the supersymmetry parameter as
\bea
\epsilon &\rightarrow & g\epsilon
\eea

\section{Our 5d SYM in the language of Ref. \cite{Hosomichi:2012ek}}\label{C}
We will now explain how to relate our hypermultiplet fields to the hypermultiplet fields in \cite{Hosomichi:2012ek}. Since the R-symmetry is broken down to $SO(4)$ or even $SO(2)\times SO(2)$ by the Scherk-Schwarz reduction, we first decompose the $SO(5)$ spinor into Weyl components of its $SO(4)$ subgroup  
\bea
\psi^{\dot\alpha} &=& \(\begin{array}{c}
\(\psi_+\)_a\\
\(\psi_-\)^r
\end{array}\)
\eea
where $a,r=+,-$. In accordance with this notation, we put the indices on our $SO(4)$ gamma matrices as 
\bea
\hat\gamma^A &=& \(\begin{array}{cc}
0 & \sigma^{A}_{as}\\
\bar\sigma^{Arb} & 0
\end{array}\)
\eea
For the $\N=1$ supersymmetry parameters we found $\epsilon^{++}$ and $\epsilon^{--}$. If we note that $\hat \Gamma^{1234} = -4s_1s_2$, then we see that these spinor components correspond to making the furher projection
\bea
\hat\Gamma^{1234} \epsilon &=& -\epsilon
\eea
This will select the Weyl components $\epsilon_a$ of the supersymmetry parameter as the ones that generate $\N=1$ supersymmetry transformations of both vector and hypermultiplets.

Vector multiplet fermions are $\(\psi_+\)_a$ and hypermultiplet fermions are $\(\psi_-\)^r$.

To conform with the notation in \cite{Hosomichi:2012ek}, which extends to cases where the number of hypermultiplets can be arbitrary, we collect our four hypermultiplet scalars into a matrix
\bea
\Phi &=& \frac{1}{2} \sum_{A=1}^4 \phi^A \sigma^A
\eea
Then we define the new hypermultiplet scalars as minus the second column in this matrix,
\bea
q_a &=& -\Phi_{a-}
\eea
Explicitly we get
\bea
q_1 &=& -\phi^1 + i \phi^2\cr
q_2 &=& \phi^3 + i \phi^4
\eea
We also define $q^a = (q_a)^*$.

For the hypermultiplet fermions which are two-component Weyl spinors $\(\psi_-\)^r$, we define a complex spinor as 
\bea
\psi_- := \(\psi_-\)^{r=+}
\eea

In order to match with the 5d Majorana condition used in \cite{Hosomichi:2012ek}, we can not use the 11d Majorana representation. Instead we choose the 11d charge conjugation matrix to be
\bea
C_{11d} &=& C_{\alpha\beta} \epsilon_{IJ} C_{\dot\alpha\dot\beta}
\eea
The 11d Majorana condition reads
\bea
\psi^{\dag} \Gamma^0 &=& \psi^T C
\eea
Writing out all spinor indices, this condition amounts to 
\bea
\(\psi^{\beta J \dot\beta}\)^* &=& \psi^{\alpha I \dot \alpha} C_{\alpha\beta} \delta_{IJ} C_{\dot\alpha\dot\beta}
\eea
We then in particular find that  
\bea
(\psi^{\alpha + r})^* &=& -\psi^{\beta + s} C_{\beta\alpha} \epsilon_{sr}
\eea
In terms of these variables, we obtain the Lagrangian that we get from (\ref{EUCL}) by $\phi^5 = i \sigma$ and $\frac{\lambda}{R} = \frac{i\mu}{R}$ as
 \bea
4\pi^2 \(-i {{\cal{L}}}^{vector}\) &=& \frac{1}{4} F^{mn} F_{mn} - \frac{1}{2}\partial^m \sigma \partial_m \sigma - \frac{2}{r^2} \sigma^2\cr
&& - \frac{i}{2}\psi_a \epsilon^{ab} \gamma^m D_m \psi_b - \frac{1}{4r} \psi_a (\sigma^3)^{ab} \psi_b\cr
4\pi^2 \(-i {{\cal{L}}}^{hyper}\) &=& \frac{1}{2}\partial_m q^a \partial^m q_a + i \psi^{\dag} \gamma^m D_m \psi + \frac{\mu}{R} \psi^{\dag} \psi\cr
&& + \frac{1}{2}\(\frac{15}{4r^2} - \(\frac{\mu}{R}\)^2\) q_a q^a - \frac{1}{2r} \frac{\mu}{R} q_a (\sigma^3)^a{}_b q^b
\eea
These Lagrangians match with the Lagrangians in \cite{Hosomichi:2012ek} for one hypermultiplet coupled to a vector multiplet, provided that we identify $\frac{\mu}{R}$ as the VEV of a vector multiplet scalar $\sigma$.

\section{Conformal mass}\label{D}
We will use the terminology that a Lagrangian of a scalar field in $d$ dimensions that is given by 
\bea
{{\cal{L}}}_{\phi} &=& -\frac{1}{2} \partial^M \phi \partial_M \phi - \frac{\R}{8} \frac{d-2}{d-1} \phi^2
\eea
describes a massless scalar field since there is only the conformal mass term. This definition applies to Euclidean and Lorentzian signature alike. In $d=6$, a massless scalar will have the Lagrangian
\bea
{{\cal{L}}}_{6d} &=& -\frac{1}{2} \partial^M \phi \partial_M \phi - \frac{\R}{10} \phi^2
\eea
and in 5d a massless scalar will have the Lagrangian
\bea
{{\cal{L}}}_{5d} &=& -\frac{1}{2} \partial^m \phi \partial_m \phi - \frac{3\R}{32} \phi^2
\eea
Time reduction of a massless scalar in 6d gives rise to a massive scalar in 5d theory, described by the Lagrangian 
\bea
{{\cal{L}}}_{5d} &=& -\frac{1}{2} \partial^m \phi \partial_m \phi - \frac{\R}{10} \phi^2\cr
&=& \[-\frac{1}{2} \partial^m \phi \partial_m \phi - \frac{3\R}{32} \phi^2\] - \frac{\R}{160} \phi^2
\eea
For the $S^5$ of radius $r$ we have $\R = \frac{20}{r^2}$ and we find that the mass term becomes
\bea
- \frac{\R}{160} \phi^2 &=& - \frac{1}{8r^2} \phi^2
\eea
so the mass of such a 5d scalar field is given by 
\bea
m_{5d} &=& \pm \frac{1}{2r}
\eea
These are precisely the critical values of the hypermultiplet mass where we find enhanced $\N=2$ supersymmetry.

\section{Unitary map between 11d and 10d gamma matrices}\label{U}
In 6d $(2,0)$ theory we impose a Weyl condition $\Gamma^{012345} \psi = \psi$. In 10d $\N=1$ SYM we impose the Weyl condition $\t\Gamma^5 \psi = \psi$ where $\t\Gamma^{012346789(10)} = \t\Gamma^5$ and these are 10d gamma matrices. Dimensional reduction of 6d theory along $x^5$ spatial direction gives 5d SYM with the same 6d Weyl projection, while dimensional reduction of 10d SYM gives the 5d SYM with the 10 Weyl projection. But these theories are isomorphic, and so there is a unitary map between the gamma matrices
\bea
\t\Gamma^M &=& U^{\dag} \Gamma^M U\cr
U &=& \frac{1}{\sqrt{2}} \(1 + \Gamma^{01234}\)
\eea
where $U^{\dag} U = 1$. This can be extended to time reduction. In this case we take $U = \frac{1}{\sqrt{2}} \(1 + i \Gamma^{12345}\)$. Then $\t\Gamma^0 \psi = i\psi$.

Given this unitary map, we can obtain allowed Lorentz groups of 10d $\N=1$ theories using either representation of the gamma matrices. Let us make the ansatz $SO(1+n,9-n)$ and pick $C = \t\Gamma^0$. Then the 10d Majorana condition $\psi^{\dag} \t\Gamma^0 \t\Gamma^{1...n} = \psi^T C$ is consistent if and only if $B^*B = 1$ with $B=\t\Gamma^{1...n}$. This is the case if $(n-1)n\in 4\mb{Z}$. But the 10d Weyl condition can only be imposed if $\t\Gamma^0$ commutes with $\t\Gamma^{1...n}$ which is if $n$ is even. This leaves us with only three cases, $n=0,4,8$ and hence the only allowed Lorentz groups are $SO(1,9)$, $SO(5,5)$ and $SO(9,1)$.

\end{document}